\documentclass[aps,prc,twocolumn,showpacs,nofootinbib]{revtex4-1}
\usepackage[utf8]{inputenc}
\usepackage{graphicx}   
\usepackage{latexsym}   
\usepackage{enumerate}
\usepackage{rotating,booktabs,multirow}
\usepackage{amsmath}
\usepackage{amsfonts}
\usepackage{amssymb}
\usepackage{multirow}
\usepackage{bm}
\usepackage{xcolor}

\begin{document}
\title{3-dimensional flow analysis: \\
A novel tool to study the collision geometry and the Equation-of-State}
\author{Tom Reichert$^{1,3}$, Jan~Steinheimer$^{4}$, Marcus~Bleicher$^{1,2,3}$}

\affiliation{$^1$ Institut f\"ur Theoretische Physik, Goethe Universit\"at Frankfurt, Max-von-Laue-Strasse 1, D-60438 Frankfurt am Main, Germany}
\affiliation{$^2$ GSI Helmholtzzentrum f\"ur Schwerionenforschung GmbH, Planckstr. 1, 64291 Darmstadt, Germany}
\affiliation{$^3$ Helmholtz Research Academy Hesse for FAIR (HFHF), GSI Helmholtz Center for Heavy Ion Physics, Campus Frankfurt, Max-von-Laue-Str. 12, 60438 Frankfurt, Germany}
\affiliation{$^4$ Frankfurt Institute for Advanced Studies (FIAS), Ruth-Moufang-Str.1, D-60438 Frankfurt am Main, Germany}

\begin{abstract}
We propose to extend the commonly known flow analysis in the transverse $p_x$-$p_y$ plane to novel flow coefficients based on the angular distribution in the $p_x$-$p_z$ and $p_y$-$p_z$ planes. The new flow coefficients, called $u_n$ and $w_n$ (in addition to $v_n$), turn out to be also highly sensitive to the nuclear Equation-of-State and can be used to explore the EoS in more detail than is possible using only $v_n$. As an example to quantify the effect of the EoS, the Ultra-relativistic Quantum Molecular Dynamics (UrQMD) model is used to investigate 20-30\% central Au+Au collisions at E$_\mathrm{lab}=1.23~A$GeV.  
\end{abstract}

\maketitle

\section{Introduction}
Quantum Chromo Dynamics (QCD), the theory of the strong interaction, is studied in heavy-ion collision experiments in todays most advanced accelerator facilities. Here, heavy atomic nuclei are accelerated to nearly the speed of light and collide head-on creating a rapidly expanding, hot and dense system consisting of hadrons and possibly subnuclear particles. Such collisions posses a remarkable resemblance with the early universe and provide an excellent opportunity to study important properties of astronomical object, e.g. compact stellar objects or binary neutron star mergers, in a controlled laboratory setting \cite{HADES:2019auv,Most:2018eaw,Most:2022wgo}. The connections between heavy-ion collisions (HIC) and astrophysics are manifold: Firstly, the nuclear Equation-of-State (EoS) is cruciual to understand the static properties of compact stars, i.e. their mass and radius, secondly, the gravitational waves emitted from binary neutron star mergers \cite{LIGOScientific:2018cki} are also sensitive to the EoS of the neutron matter. Thirdly, the magnetic fields generated during a heavy ion collision are similar to the ones expected in neutron stars and finally also the vorticities present in HICs and in stellar objects are on the same order of magnitude \cite{STAR:2017ckg,Becattini:2015ska,Becattini:2021iol}. Especially, the precise determination of the nuclear equation-of-state of the QCD-matter at high baryon densities is currently a major topic of low and intermediate energy heavy ion reactions. Here, the currently running High Acceptance Di-Electron Spectrometer (HADES) experiment at GSI \cite{HADES:2009aat} has already shown its capabilities to yield substantial new information on the EoS of super dense baryonic matter \cite{Kardan:2018hna,HADES:2020lob}. 

A central tool to extract the EoS of QCD-matter has been the investigation of the expansion flow of nucleons in terms of the Fourier decomposition of the angular emission in the transverse plane (i.e. the $p_x$-$p_y$ plane). These coefficients are usually denoted as $v_n$, where the lowest coefficients are called the directed flow ($v_1$), the elliptic flow ($v_2$), the triangular flow ($v_3$) and the quadrangular flow ($v_4$). The measurement of even higher flow components is possible and has e.g. been done in \cite{Tang:2004je,HADES:2020lob}. A systematic comparison of the flow coefficients $v_n$ with theoretical calculations does then allow to extract the properties of the EoS, see e.g. \cite{Danielewicz:2002pu,Hillmann:2018nmd,Hillmann:2019wlt,Mohs:2020awg}.

In this paper we extend this well known tool to include also the $p_x$-$p_z$ and the $p_y$-$p_z$ plane. We call these novel flow coefficients $u_n$ ($p_x$-$p_z$ plane) and $w_n$ ($p_y$-$p_z$ plane). Especially at low energies where the expansion of the system is rather isotropic due to the low energy of the incoming beam, these coefficients provide further information on the expansion pattern and the correlations of the created matter which can be used to infer the EoS and/or to benchmark model simulations. The novel coefficients might further be utilized to study so called anti-flow \cite{Brachmann:1999xt,Bozek:2010bi,Nara:2021fuu}.

\section{Model setup and flow extraction}
For the present analysis the Ultra-relativistic Quantum Molecular Dynamics (UrQMD 3.5) model \cite{Bass:1998ca,Bleicher:1999xi,Bleicher:2022kcu} is used. UrQMD describes the dynamics of the colliding hadronic system by solving the equations of motion for each individual hadron and includes a collision term, if the distance between two hadron becomes closer than their interaction distance defined by the total cross section.  In its current version, UrQMD contains a broad range of hadronic degrees of freedom up to masses of 4~GeV. The propagation of the hadrons can be done either in cascade mode (i.e. without potential interactions) or one can employ hadronic potentials corresponding to different equations-of-state (usually a hard and soft EoS, where hard/soft is defined by the compressibility of the nuclear matter). First predictions for bulk observables at 1.58~$A$GeV kinetic beam energy have been recently published in Ref. \cite{Reichert:2021ljd}. UrQMD has been further highly successful in describing and predicting directed, elliptic, triangular as well as quadrangular flow of nucleons and light clusters \cite{Hillmann:2018nmd,Hillmann:2019wlt} and their correlations \cite{Reichert:2022gqe}. 

The coordinate system typically used in heavy ion collisions is defined with the $z$-axis along the beam direction (for fixed target experiments positive $z$ is in the direction of the projectile), the direction of the impact parameter is usually denoted as $x$-direction and the direction orthogonal (in a right handed coordinate system) is defined as the $y$-direction. The plane spanned by $x-z$ is called the reaction plane (RP), the plane defined by $x-y$ is called the transverse plane (TP) and the remaining plane defined by $y-z$ is called the perpendicular plane (PP). In an experimental setup the reaction plane, transverse plane and perpendicular plane are unknown and need to be reconstructed in a meaningful way by estimating the event-by-event reaction plane angle $\Psi_{RP}$. The HADES experiment uses a forward wall to reconstruct the first order event plane from the spectators which is a good proxy for the reaction plane \cite{Kardan:2017knj,HADES:2020lob}. In contrast, in the simulation the coordinate system is fixed (and hence the RP, TP and PP) are known. Thus, $\Psi_{RP}=0$ in the model, also for the new flow coefficients $u$ and $w$ introduced below we will assume that the angular distribution starts strictly on the coordinate axis. Let us stress, that one can of course perform the same analysis with respect to planes defined by experimentally measured $Q$-vectors if desired. 

The well known harmonic flow coefficients $v_n$ arise from the Fourier decomposition of the angular distribution of the nucleons in the transverse plane, cf. Eq. \eqref{eq:dNdphi_v}. 
\begin{equation}\label{eq:dNdphi_v}
    \frac{\mathrm{d}N}{\mathrm{d}\phi_{xy}} = 1 + 2\sum\limits_{n=1}^\infty v_n^{\rm sym}\cos(n\phi_{xy}) + v_n^{\rm asym}\sin(n\phi_{xy})
\end{equation}
The complete decomposition includes both the symmetric cosine and the antisymmetric sine terms\footnote{The coefficients of the cosine terms will be denoted with ``sym" while the coefficients of the antisymmetric terms will be denoted by ``asym" for the $v_n$, $u_n$ and $w_n$ harmonics.}. Obviously the antisymmetric $v_n$ coefficients will be zero if averaged over events, although they might be nonzero for single events. The sine term is explicitly included in the equation because it will be an important term for the extraction of the novel $u_n$ and $w_n$ coefficients. 

In the low energy (high baryon density) regime investigated in this article the transverse flow harmonics are mostly extracted with respect to the first order event plane defined by the spectators. We point out that this extraction of the usual flow coefficients $v_n$ with respect to the reaction plane (first order event plane) is quite different from very high energies. At very high energies (e.g. at RHIC or LHC), the flow harmonics are typically measured with respect to the n-th order event plane which can be reconstructed via the flow vector $Q$ \cite{Danielewicz:1985hn,Poskanzer:1998yz,Borghini:2000sa}. Here, the competition between the flow and the statistical fluctuations has to be taken into account by a resolution parameter $\chi_n$ \cite{Ollitrault:1997di}. However, this extraction method relies on the idea that every particle correlation is due to flow. To also account for nonflow effects (e.g. quantum statistics effects, resonance decays, momentum conservation, (mini) jets, strong and Coulomb interaction, ...) \cite{Borghini:2002mv,Cheng:2000tk} it is convenient to extract the flow signal from the the 2- and 4-particle cumulants \cite{Borghini:2001vi} or from the Lee-Yang zeroes method \cite{Bhalerao:2003xf}. However, at low energies investigated in this article the flow harmonics are rather small, the multiplicities are limited and the sign of the elliptic flow is important because of the squeeze-out effect due to spectator blocking. In line with the experimental method at low energies, we fix $\Psi_{RP}=0$ in the simulation (also for the new flow coefficients we use $\Psi_{TP}=\Psi_{PP}=0$). The flow coefficients in the transverse plane are then extracted from the simulation by
\begin{align}
    &v_n^{\rm sym} = \langle \cos(n\phi_{xy}) \rangle \\
    &v_n^{\rm asym} = \langle \sin(n\phi_{xy}) \rangle,
\end{align}
in which $\tan(\phi_{xy})=p_y/p_x$ and where the average $\langle\cdot\rangle$ is taken over all participating nucleons which are not bound in light clusters in a given event.

We extend this analysis to the two other directions and introduce novel flow coefficients which are extracted from the angular distributions in the $p_x$-$p_z$ plane and the $p_y$-$p_z$ plane. The distributions can then be expressed as Fourier series, see Eqs. \eqref{eq:dNdphi_u} and \eqref{eq:dNdphi_w}.
\begin{align}\label{eq:dNdphi_u}
    \frac{\mathrm{d}N}{\mathrm{d}\phi_{xz}} &= 1 + 2\sum\limits_{n=1}^\infty u^{\rm sym}_n \cos(\phi_{xz}) + u^{\rm asym}_n \sin(\phi_{xz}) \\
    \frac{\mathrm{d}N}{\mathrm{d}\phi_{yz}} &= 1 + 2\sum\limits_{n=1}^\infty w^{\rm sym}_n \cos(\phi_{yz}) + w^{\rm asym}_n \sin(\phi_{yz}) \label{eq:dNdphi_w}
\end{align}
Here the angle $\phi_{xz}$ ($\phi_{yz}$) are defined with respect to the $x$-axis ($y$-axis). 

This Fourier decomposition introduces novel Fourier coefficients which we call $u_n$ and $w_n$. They are calculated similarly to the common $v_n$ harmonics:
\begin{align}
    &u^{\rm sym}_n = \langle \cos(n\phi_{xz}) \rangle \\
    &u^{\rm asym}_n = \langle \sin(n\phi_{xz}) \rangle \\
    &w^{\rm sym}_n = \langle \cos(n\phi_{yz}) \rangle \\
    &w^{\rm asym}_n = \langle \sin(n\phi_{yz}) \rangle
\end{align}
The interpretation of these novel coefficients is easier if they are expressed in the actual momentum variables, see Eqs. \eqref{eq:u1_mom} to \eqref{eq:w2_mom}.
\begin{align}
    &u_1^{\rm sym} = \frac{p_x}{\sqrt{p_x^2+p_z^2}}\,, \quad &&u_1^{\rm asym} = \frac{p_z}{\sqrt{p_x^2+p_z^2}} \label{eq:u1_mom} \\
    &u_2^{\rm sym} = \frac{p_x^2-p_z^2}{p_x^2+p_z^2}\,, \quad &&u_2^{\rm asym} = \frac{2p_xp_z}{p_x^2+p_z^2} \\
    &w_1^{\rm sym} = \frac{p_y}{\sqrt{p_y^2+p_z^2}}\,, \quad &&w_1^{\rm asym} = \frac{p_z}{\sqrt{p_y^2+p_z^2}} \\
    &w_2^{\rm sym} = \frac{p_y^2-p_z^2}{p_y^2+p_z^2}\,, \quad &&w_2^{\rm asym} = \frac{2p_yp_z}{p_y^2+p_z^2} \label{eq:w2_mom}
\end{align}

\section{Results: Momentum dependence of the novel flow coefficients}
The results presented here are for semi-peripheral (20-30\% centrality, corresponding to an impact parameter range of $6.6\leq b\leq8.1$ fm, cf. Ref. \cite{HADES:2017def}) Au+Au collisions at a kinetic beam energy of 1.23 $A$GeV. We compare the flow of the participating free nucleons (i.e. without spectators and excluding those bound in light clusters) from simulations employing a hard Skyrme type EoS, a soft Skyrme type EoS and in cascade mode (i.e. without potential interactions). 

\begin{figure} [t!hb]
    \centering
    \includegraphics[width=\columnwidth]{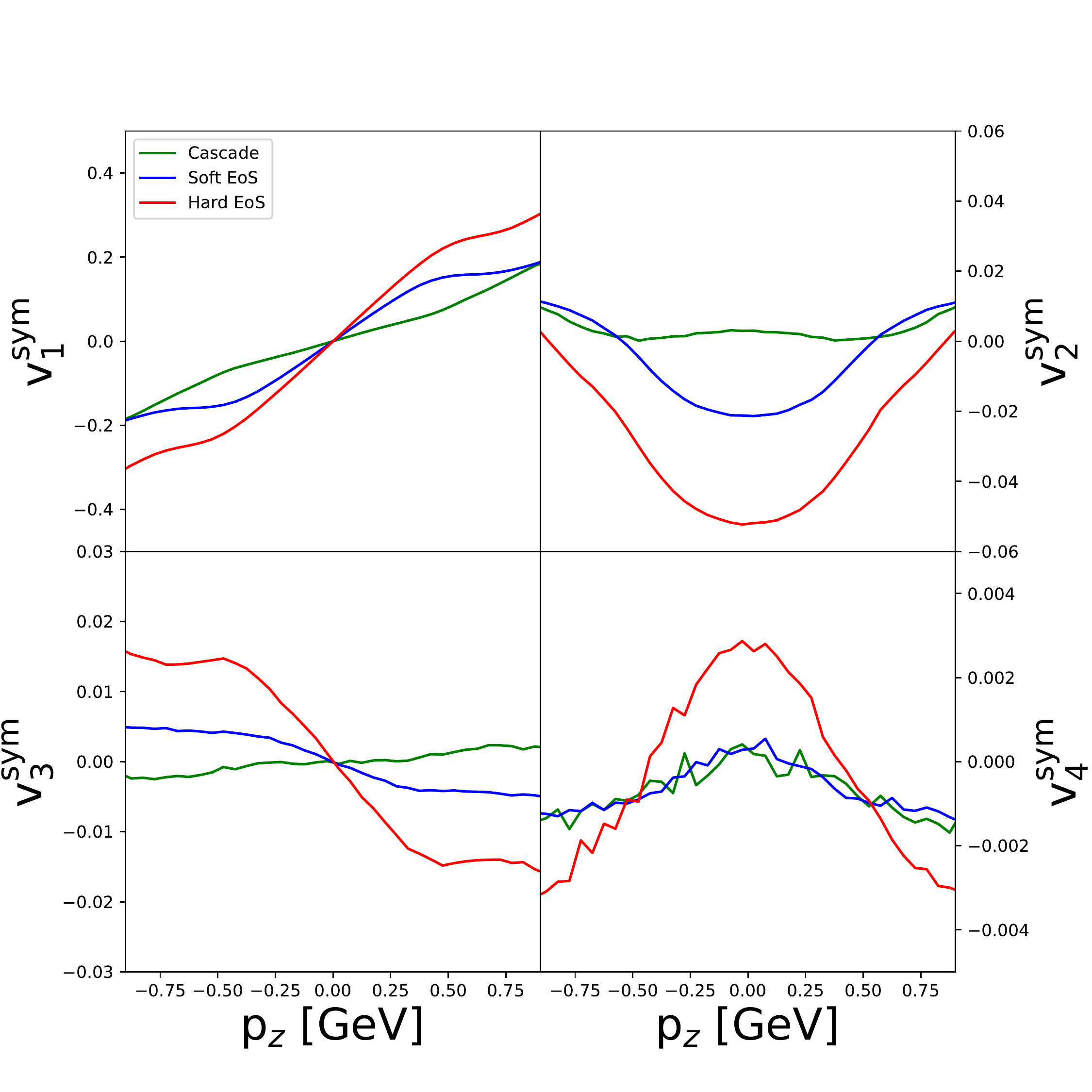}
    \caption{[Color online] The first four symmetric flow coefficients $v_1^{\rm sym}$ (upper left), $v_2^{\rm sym}$ (upper right), $v_3^{\rm sym}$ (lower left), $v_4^{\rm sym}$ (lower right) in dependence of $p_z$ from 20-30\% central Au+Au collisions at E$_\mathrm{lab}=1.23$~$A$GeV from UrQMD with a hard EoS (red), a soft EoS (blue) and in cascade mode (green).}
    \label{fig:vn_pz}
\end{figure}

We start our investigation with the well known flow coefficients $v_n$ of the angular distribution in the transverse plane. Here, Fig. \ref{fig:vn_pz} shows the first four symmetric flow coefficients $v_1^{\rm sym}$ (upper left), $v_2^{\rm sym}$ (upper right), $v_3^{\rm sym}$ (lower left), $v_4^{\rm sym}$ (lower right) as a function of the longitudinal momentum $p_z$. The antisymmetric $v_n^{\rm asym}$ average to zero due to symmetry and because the reaction plane is fixed in the simulation.
The directed flow $v_1^{\rm sym}$ has a positive correlation with the longitudinal momentum and reflects the strength of the bounce off of the two incoming nuclei as was discussed in detail in \cite{Reichert:2022gqe} and we observe as expected that with a stiffer equation of state the bounce off becomes more pronounced due to larger pressure (density) gradients. The elliptic flow $v_2^{\rm sym}$ is negative around zero longitudinal momenta and exhibits a convex shape. A stiffer EoS leads to a stronger squeeze out effect and  the second flow coefficients become more negative. The triangular flow coefficient $v_3^{\rm sym}$ has a negative correlation with the longitudinal momentum for the hard and soft EoS, again with the stiffest EoS showing the largest magnitude, while the cascade mode has a positive correlation but its value is close to zero. The triangular flow is in this low energy and high density regime often connected to the direct product between the directed and the elliptic flow, i.e. $v_3^{\rm sym}=v_1^{\rm sym}\cdot v_2^{\rm sym}$. Lastly, the quadrangular flow $v_4^{\rm sym}$ is positive around zero longitudinal momenta and exhibits a concave shape. Its magnitude decreases with decreasing stiffness of the EoS.  

\begin{figure} [t!hb]
    \centering
    \includegraphics[width=\columnwidth]{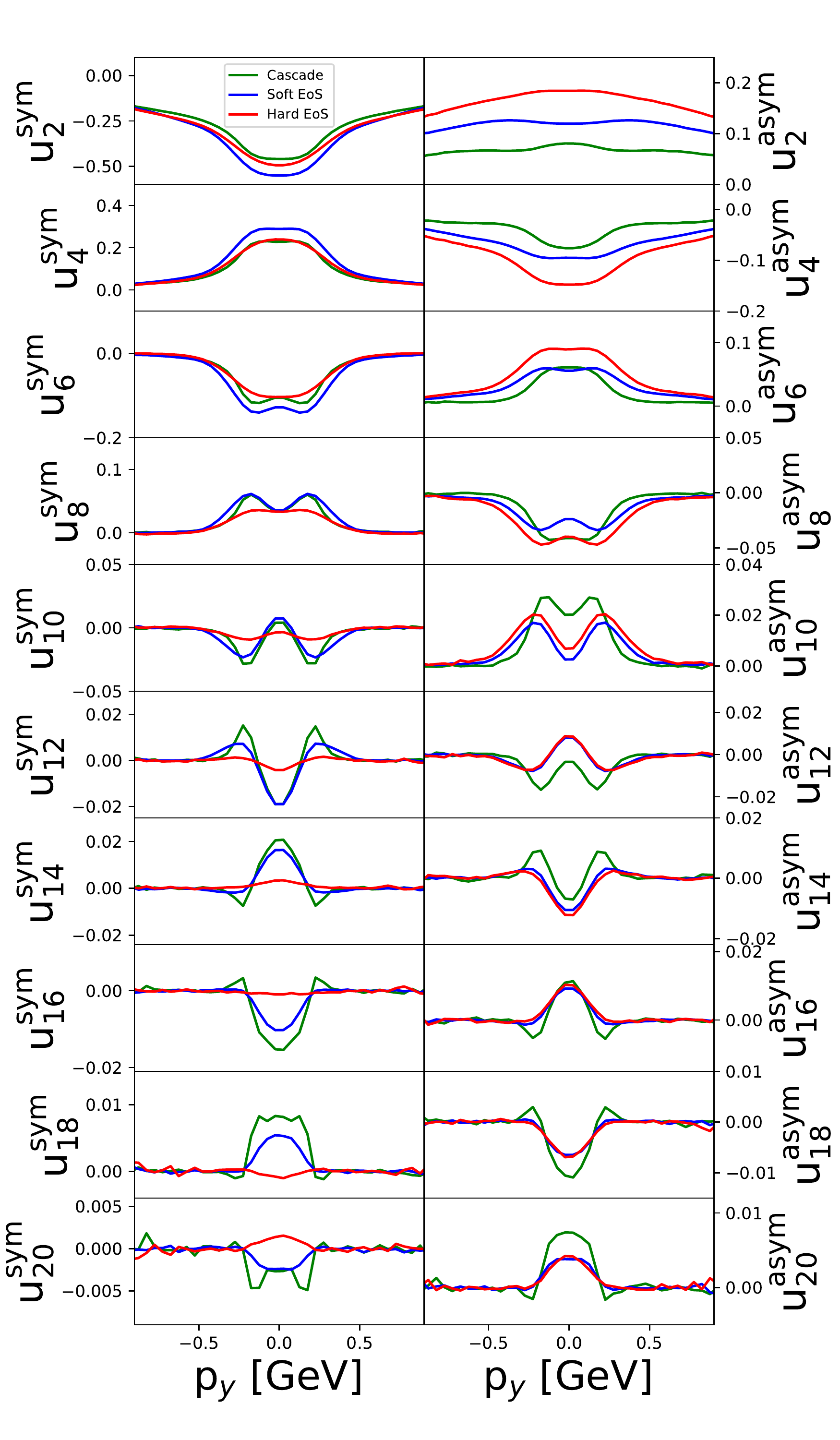}
    \caption{[Color online] The first twenty non-zero symmetric flow coefficients $u_n^{\rm sym}$ (left column) and the antisymmetric coefficients $u_n^{\rm asym}$ (right column) in dependence of $p_y$ from 20-30\% central Au+Au collisions at E$_\mathrm{lab}=1.23$~$A$GeV from UrQMD with a hard EoS (red), a soft EoS (blue) and in cascade mode (green).}
    \label{fig:un_py}
\end{figure}

After setting the stage, we turn now to the newly introduced flow coefficients extracted from the other angles, starting with the $u_n$ coefficients (using the angle in the reaction plane, i.e. the $p_x$-$p_z$ plane). Here, the odd flow coefficients (i.e. the coefficients whose order is odd) average to zero due to symmetry and the fixed beam direction. The twenty lowest non-vanishing $u_n$ coefficients are shown in Fig. \ref{fig:un_py}. From top to bottom we depict the first to the twentieth symmetric flow coefficients $u_n^{\rm sym}$ (left column) and the antisymmetric coefficients $u_n^{\rm asym}$ (right column) in dependence of $p_y$ for 20-30\% central Au+Au collisions at E$_\mathrm{lab}=1.23$~$A$GeV from UrQMD with a hard EoS (red), a soft EoS (blue) and in cascade mode (green).
In contrast to the usual $v_n$ coefficients whose higher components ($n>4$) are very difficult to measure and to calculate due to the nearly spherical shape of the transverse expansion, the new $u_n$ coefficients reveal a large signal up to the 20th order. The main reason for the largeness of these coefficients is the large average longitudinal momentum as compared to the transverse momenta leading to strong asymmetries in the transverse vs. longitudinal planes. This strong asymmetry can also be seen in the angular distributions, discussed in the following section.

Before turning to the angular distributions, let us investigate the coefficients step by step beginning with the symmetric $u_n^{\rm sym}$ in the left column. As expected all coefficients posses a mirror symmetry around $p_y=0$. The symmetric even coefficients extracted from simulations with a hard EoS whose order can be written as $4n+2$ have an overall negative sign, while those which can be written as $4n$ posses an overall positive sign. Their shape is consequently alternating between convex and concave around zero $p_y$. But most prominent is the finding that the results obtained from a hard EoS show only one local maximum or minimum. This situation is different for the soft EoS and for the simulation in cascade mode. With decreasing stiffness, the even symmetric flow coefficients develop local structures counteracting the overall trend of the curves more drastically. The local extreme at zero $p_y$ in the cascade simulations even invert the alternating sign pattern at n=10. Finally, it should be noted that the coefficients from both simulations with Skyrme potential level off more rapidly with increasing order than the cascade simulations. This dependence on the equation of state is not surprising as not only the transverse dynamics are influenced by the EoS, but also the relation between longitudinal and transverse dynamics. Thus, the whole 3-dimensional shape of the momentum distributions is influenced by the EoS which is then reflected in the expansion coefficients.

Now let us turn to the right column of Fig. \ref{fig:un_py} showing the antisymmetric even $u_n$ coefficients. The alternating sign pattern of the coefficients can also be observed here, but opposite to the pattern of the symmetric coefficients. Here, the hard EoS has the largest magnitude among the different simulation modes until the 8th order, after which the cascade mode simulations show the largest magnitude. With increasing order more structure appears again around $p_y=0$. To summarize the investigation of the non-vanishing $u_n$ coefficients extracted in the $p_x$-$p_z$ plane, we observe that: 1.) the coefficients can be extracted to much higher order and still show a significant signal, 2.) the $u_n$ coefficients show detailed structures with increasing order and 3.) the different employed equations of state produce flow patterns which are highly different allowing to measure the nuclear EoS more precisely. 

\begin{figure} [t!hb]
    \centering
    \includegraphics[width=\columnwidth]{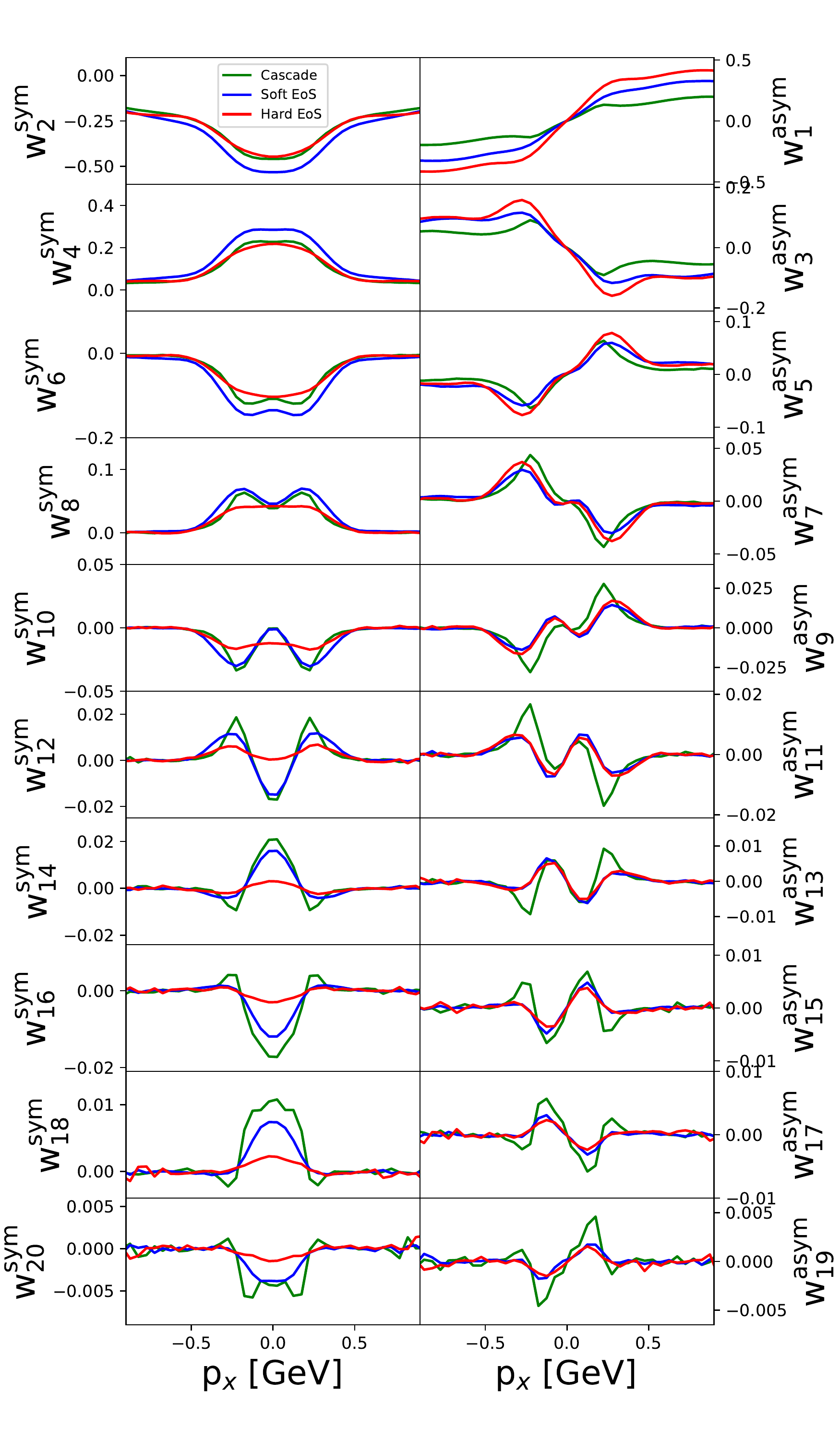}
    \caption{[Color online] The first twenty non-zero symmetric even flow coefficients $w_n^{\rm sym}$ (left column) and the antisymmetric odd $u_n^{\rm asym}$ (right column) in dependence of $p_x$ from 20-30\% central Au+Au collisions at E$_\mathrm{lab}=1.23$~$A$GeV from UrQMD with a hard EoS (red), a soft EoS (blue) and in cascade mode (green).}
    \label{fig:wn_px}
\end{figure}

Finally, we turn to the investigation of the $w_n$ coefficients (i.e. the angular distribution in the $p_y$-$p_z$ plane). Here, the symmetric odd coefficients and the antisymmetric even coefficients average to zero due to symmetry and the fixed perpendicular plane. Fig. \ref{fig:wn_px} shows the first twenty non-zero symmetric even flow coefficients $w_n^{\rm sym}$ (left column) and the antisymmetric odd $u_n^{\rm asym}$ (right column) in dependence of $p_x$ from 20-30\% central Au+Au collisions at E$_\mathrm{lab}=1.23$~$A$GeV from UrQMD with a hard EoS (red), a soft EoS (blue) and in cascade mode (green).
Let us begin the discussion of the symmetric $w_n$ flow harmonics in the left column of Fig. \ref{fig:wn_px}. The symmetric $w_n$ coefficients for the cascade simulations are very similar in structure and magnitude to the symmetric $u_n$ coefficients from the cascade simulations. Differences between the $w_n^{\rm sym}$ and the $u_n^{\rm sym}$ only become visible for the higher order harmonics in case of the simulations with Skyrme type potential. The $w_n$ with a hard EoS produce a slightly flatter structure than the $u_n$. Despite the similarities between the symmetric $u_n$ and $w_n$ flow coefficients, the antisymmetric $w_n$ coefficients shown in the right column of Fig. \ref{fig:wn_px} are very different. In case of collisions with equal target and projectile mass number, they are point symmetric around $p_x=0$ by definition. For the first five antisymmetric $w_n$ coefficients, the simulations show a positive slope at zero $p_x$ for all coefficients whose order can be represented as $4n+1$ and a negative slope for the coefficients with order $4n+3$. However, the curves are again developing local structures changing the slope around $p_x=0$ after the seventh order. For the higher orders, the coefficients extracted from simulations with larger stiffness level off more rapidly. We conclude that also the $w_n$ coefficients are rich in structure and are highly sensitive probes of the nuclear equation of state and their investigation will allow to pin down the expansion geometry in greater detail than a restriction to the known $v_n$ coefficients alone.

\subsection{Angular distributions}
We continue our exploration with the angular distribution of the participating free nucleons. The distributions are obtained by calculating the average symmetric and antisymmetric flow harmonics up to 20th order and inserting them into Eqs. \eqref{eq:dNdphi_v}, \eqref{eq:dNdphi_u} and \eqref{eq:dNdphi_w}. All polar plots in this section are normalized to $2\pi$.

\begin{figure} [t!hb]
    \centering
    \includegraphics[width=\columnwidth]{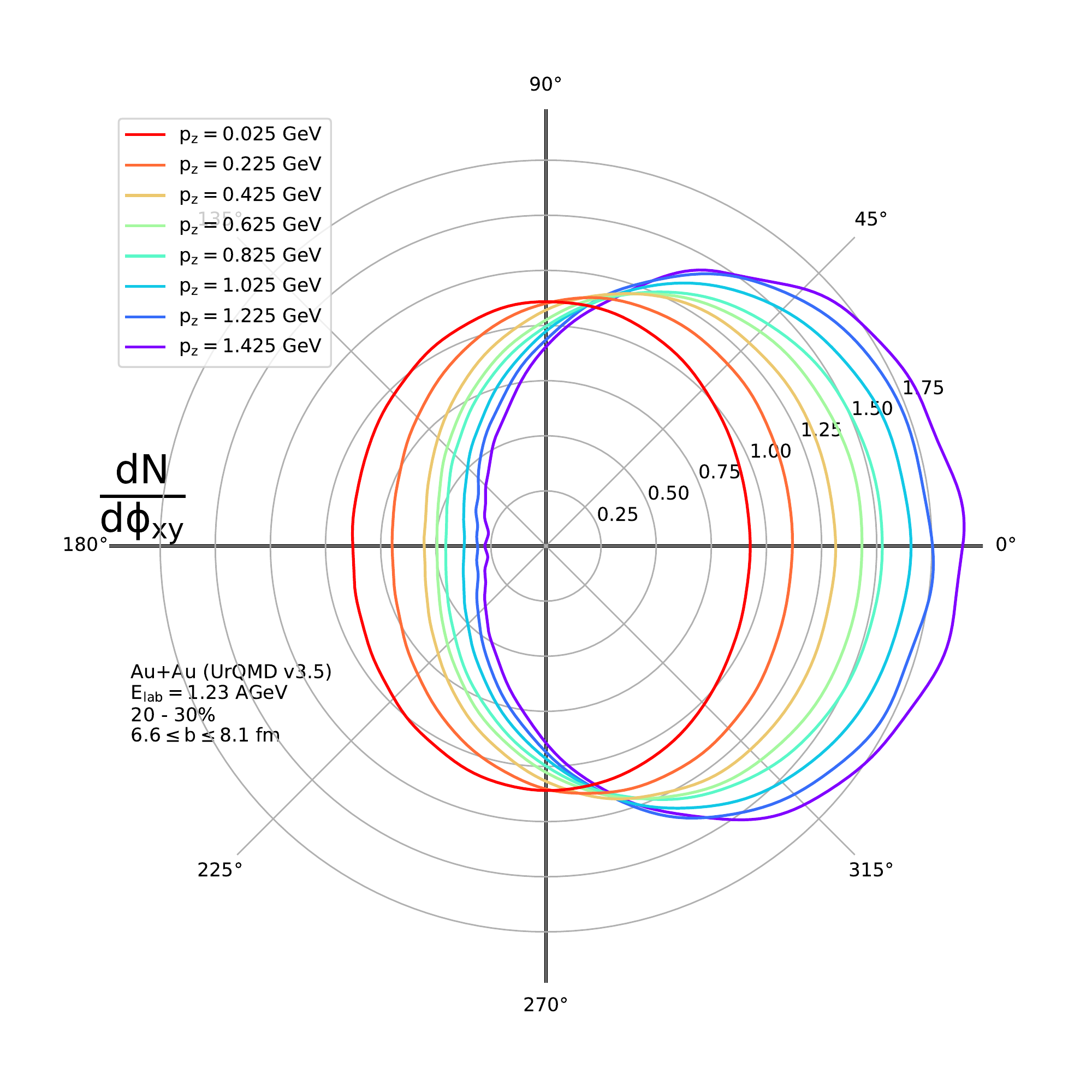}
    \caption{[Color online] The angular distribution of participating free nucleons in the $p_x$-$p_y$ plane in different $p_z$ bins from 20-30\% central Au+Au collisions at E$_\mathrm{lab}=1.23$~$A$GeV from UrQMD with a hard EoS. Each curve is normalized to $2\pi$.}
    \label{fig:polar_v}
\end{figure}

First, Fig. \ref{fig:polar_v} shows the angular distribution of participating free nucleons in the $p_x$-$p_y$ plane in different $p_z$ bins from 20-30\% central Au+Au collisions at E$_\mathrm{lab}=1.23$~$A$GeV from UrQMD with a hard EoS. 
The plot reveals interesting features: Firstly, at zero momentum in beam direction, an elliptic shape can be observed whose large semi-axis is pointing out-of plane as expected in this energy regime. This is due to the spectator shadowing and the space and time dependent emission pattern and is further lining up with the measured data \cite{HADES:2020lob}. Secondly, one can observe that at large (positive) longitudinal momenta, the shape of the transverse momentum space is becoming more circular while additionally receiving a shift towards positive $p_x$. This reflects the bounce off of the impinging nuclei.

\begin{figure} [t!hb]
    \centering
    \includegraphics[width=\columnwidth]{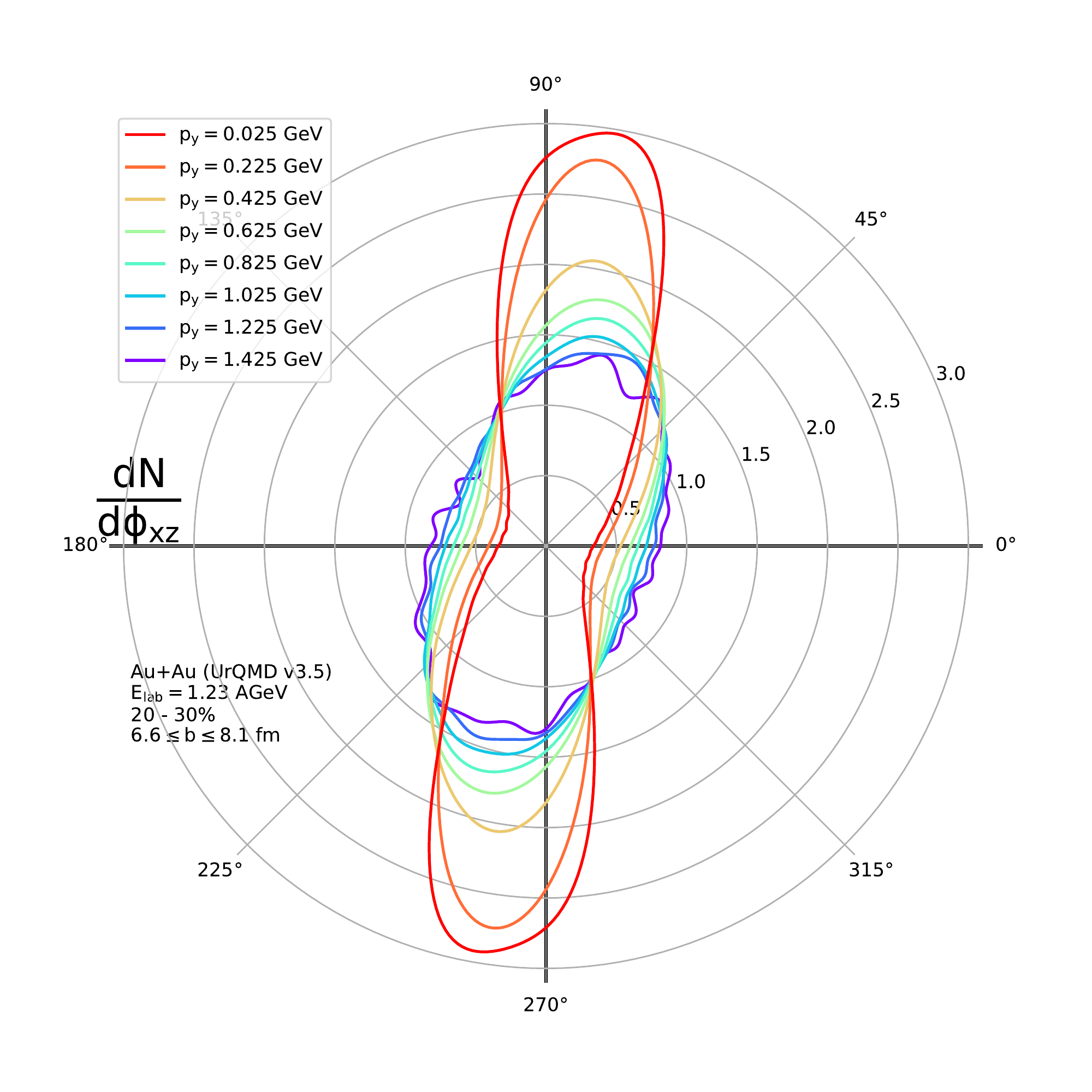}
    \caption{[Color online] The angular distribution of participating free nucleons in the $p_x$-$p_z$ plane in different $p_y$ bins from 20-30\% central Au+Au collisions at E$_\mathrm{lab}=1.23$~$A$GeV from UrQMD with a hard EoS. Each curve is normalized to $2\pi$.}
    \label{fig:polar_u}
\end{figure}

Next we turn to the discussion of the momentum space distribution coming from the novel $u_n$ coefficients. Fig. \ref{fig:polar_u} shows the angular distribution of participating free nucleons in the $p_x$-$p_z$ plane in different $p_y$ bins from 20-30\% central Au+Au collisions at E$_\mathrm{lab}=1.23$~$A$GeV from UrQMD with a hard EoS.
In contrast to the transverse angular distribution where the antisymmetric contributions average to zero, here the situation is clearly different. For $p_y=0$ the distribution is highly forward-backward peaked with a slight tilt reflecting the bounce off. It is however interesting that with increasing $p_y>0$ the direction of the bounce off is rotating farther away from the beam direction, but also the overall asymmetry of the distribution is strongly decreased. This shows the strong effect of spectator shadowing in the reaction plane and its decrease as one moves out of the reaction plane (towards large $p_y$).

\begin{figure} [t!]
    \centering
    \includegraphics[width=\columnwidth]{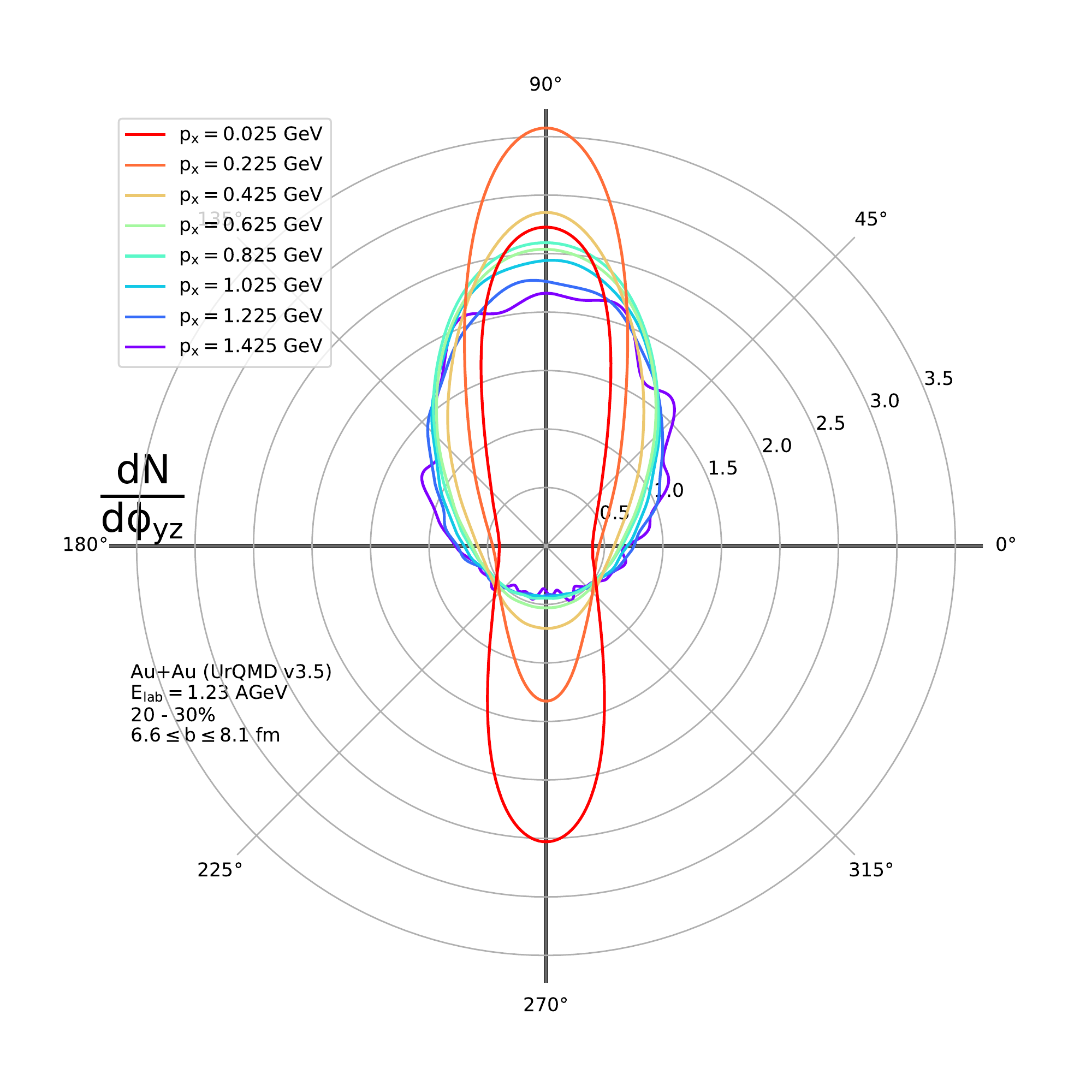}
    \caption{[Color online] The angular distribution of participating free nucleons in the $p_y$-$p_z$ plane in different $p_x$ bins from 20-30\% central Au+Au collisions at E$_\mathrm{lab}=1.23$~$A$GeV from UrQMD with a hard EoS. Each curve is normalized to $2\pi$.}
    \label{fig:polar_w}
\end{figure}

Lastly, we investigate the angular distribution evaluated from the novel $w_n$ flow coefficients. Fig. \ref{fig:polar_w} shows the angular distribution of participating free nucleons in the $p_y$-$p_z$ plane in different $p_x$ bins from 20-30\% central Au+Au collisions at E$_\mathrm{lab}=1.23$~$A$GeV from UrQMD with a hard EoS.
Here we observe the momentum space distribution from the side and only for positive $p_x$. Therefore, all the distributions are obviously peaked towards positive $p_z$ because of the bounce-off, positive $p_x$ values select mainly projectile particles. It can be further noticed that at $p_x=0$ the distribution is strongly forward peaked with very little contribution in $p_y$ direction, while with increasing $p_x$ the distribution is becoming more spherical. This finding is mainly due to four momentum conservation, i.e. increasing stopping allows for more transverse expansion.

\subsection{Impact on the bounce off}
\begin{figure} [t!]
    \centering
    \includegraphics[width=\columnwidth]{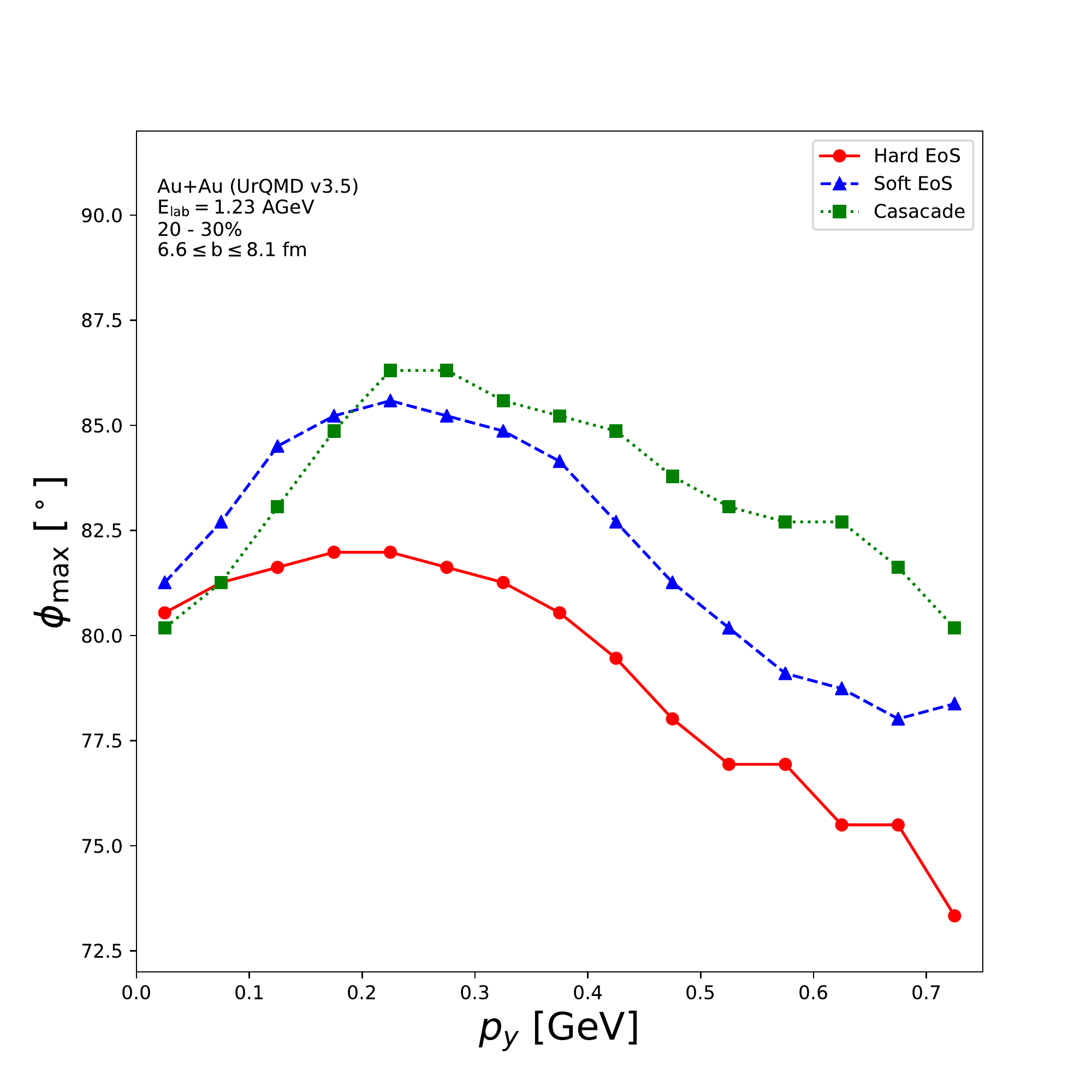}
    \caption{[Color online] The tilt of the peak of the angular distribution of participating free nucleons in the $p_x$-$p_z$ plane as a function of $p_y$ from 20-30\% central Au+Au collisions at E$_\mathrm{lab}=1.23$~$A$GeV from UrQMD with a hard EoS (red cirlces), a soft EoS (blue triangles) and in cascade mode (green squares).}
    \label{fig:rot_u}
\end{figure}

The investigation of the $u_n$ flow harmonics and the respective reconstructed angular distribution suggests that the direction of the bounce-off tilts with increasing out-of plane momentum ($p_y$). To quantify this in more detail, we investigate the angle of the peak of the respective distribution as a function of the $p_y$ momentum. The peak in the first quadrant (positive $p_x$, positive $p_z$) is selected for this. Fig. \ref{fig:rot_u} shows the tilt of the peak of the angular distribution of participating free nucleons in the $p_x$-$p_z$ plane as a function of $p_y$ from 20-30\% central Au+Au collisions at E$_\mathrm{lab}=1.23$~$A$GeV from UrQMD with a hard EoS (red cirlces), a soft EoS (blue triangles) and in cascade mode (green squares).
When comparing the simulations with the Skyrme type potential and without potential, it can be firstly observed that the peaks for all equations of state are located at $p_y=0.2$ GeV and rotate continuously towards smaller angles for smaller and larger $p_y$, i.e. towards spherical expansion. While at $p_y=0$ all EoS show the same rotation, it can be observed that the hard EoS receives a stronger rotation of the peak angle (reflected in a larger shift to smaller angles) at $p_y>0$, i.e. a stronger bounce off, which is in line with the expectation. Due to its larger incompressibility the hard EoS leads to a higher pressure increasing the expansion. With decreasing strength of the nuclear potential, binary scattering reactions as well as resonance formation and decays become more important suppressing the generation of significant elliptic flow $v_2$ which indicates that the transverse expansion is getting isotropic. It is interesting to note that such a rotation of the emission source can also bee seen in the HBT analysis \cite{Lisa:2011na} and the dependence of the rotation angle on $p_y$ strongly resembles the 'cat eye' structure predicted for HBT correlations in \cite{Graef:2013wta}.

\section{Conclusion}
We have employed the Ultra-relativistic Quantum Molecular Dynamics model (UrQMD) to simulate 20-30\% semi-peripheral Au+Au collisions at E$_{\rm lab}=1.23$ $A$GeV with different equations-of-state. Due to the complex 3-dimensional collision dynamics, additional information on the equation of state can be gained by investigating not only the momentum asymmetries in the transverse plane but also by including additional symmetry planes. In this work we introduced the novel flow coefficients $u_n$ extracted from the angular distribution in the $p_x$-$p_z$ plane and $w_n$ extracted from the angular distribution in the $p_y$-$p_z$ plane. These novel flow coefficients have significant signals up to high order and show a high sensitivity to the employed equation-of-state. In contrast to the commonly used $v_n$ coefficients connected to the angular distribution in the transverse plane which are strongly damped for $n>6$, mostly show quantitative differences among EoS with varying stiffness, the novel $u_n$ and $w_n$ flow harmonics are rich in structure and show also large qualitative differences for higher orders. Their measurement will allow to constrain the nuclear equation-of-state more precisely than is possible with $v_n$ alone. 

\begin{acknowledgements}
The authors thank Behruz Kardan and Christoph Blume for fruitful discussion about the flow harmonics and the analysis. This article is part of a project that has received funding from the European Union’s Horizon 2020 research and innovation programme under grant agreement STRONG – 2020 - No 824093. J.S. thanks the Samson AG for funding. Computational resources were provided by the Center for Scientific Computing (CSC) of the Goethe University and the ``Green Cube" at GSI, Darmstadt. This project was supported by the DAAD (PPP Thailand). This research has received funding support from the NSRF via the Program Management Unit for Human Resources \& Institutional Development, Research and Innovation [grant number B16F640076].
\end{acknowledgements}



\end{document}